\documentclass[twocolumn,showpacs,aps,prl]{revtex4}

\usepackage{times,amsmath,amsfonts,amssymb,bbm,graphicx}

\newcommand{\gr}[1]{\boldsymbol{#1}}

\newcommand{\sig}{\gr{\sigma}}
\newcommand{\eq}[1]{Eq.~(\ref{#1})}
\newcommand{\ineq}[1]{Ineq.~(\ref{#1})}

\begin{document}

\title{Strong Monogamy of Bipartite and Genuine Multipartite Entanglement: The Gaussian Case}

\author{Gerardo Adesso$^{1,2}$ and Fabrizio Illuminati$^{2,3,4}$}
\affiliation{$^{1}$Dipartimento di Fisica ``E. R. Caianiello'',
Universit\`{a} degli Studi di
Salerno, Via S. Allende, 84081 Baronissi (SA), Italy. \\
$^{2}$CNR-INFM Coherentia, Unit\`{a} di Salerno, CNISM, and INFN Sezione di Napoli-Gruppo Collegato di Salerno. \\
$^{3}$Dipartimento di Matematica e Informatica, Universit\`{a} degli
Studi di Salerno, Via Ponte Don Melillo, 84084 Fisciano (SA),
Italy. \\
$^{4}$ISI Foundation for Scientific Interchange, Viale S. Severo 65,
10133 Torino, Italy.}

\date{August 17, 2007}

\begin{abstract}
We demonstrate the existence of general constraints on distributed
quantum correlations, which impose a trade-off on  bipartite {\em
and}
 multipartite entanglement at once. For all $N$-mode Gaussian
states under permutation invariance, we establish exactly a monogamy
inequality, stronger than the traditional one, that by recursion
defines a proper measure of genuine $N$-partite entanglement. Strong
monogamy holds as well for subsystems of arbitrary size, and the
emerging multipartite entanglement measure is found to be scale
invariant. We unveil its operational connection with  the optimal
fidelity of continuous variable teleportation networks.
\end{abstract}
\pacs{03.67.Mn, 03.65.Ud}
\maketitle
Understanding the structure of  entanglement distributed among many
parties is central to diverse aspects of quantum information theory
\cite{hororev} and its manifold applications in condensed matter
physics \cite{faziorev}.  A direct consequence of the no-cloning
theorem \cite{nocloning} is what one might call the {\em no-sharing}
theorem: maximal entanglement cannot be freely shared. Suppose Alice
is maximally entangled to both Bob and Charlie, then she could
exploit both channels to teleport two perfect clones of an unknown
state,  violating the linearity of quantum mechanics. Nonmaximal
entanglement, however, can be shared; but this distribution is
constrained to {\em monogamy} inequalities \cite{pisa}, as
originally discovered by Coffman, Kundu, and Wootters (CKW)
\cite{CKW}. In the most general known form, monogamy imposes the
following trade-off on bipartite entanglement distributed among $N$
parties $p_1 \ldots p_N$,
\begin{equation}
\label{ckwine} E^{p_1 \vert (p_2 \ldots p_N
)} \ge \begin{array}{c}\sum_{j \ne 1}^N {E^{p_1 \vert p_j }
}\end{array}\, ,
\end{equation}
where  $E$ is a proper measure of bipartite entanglement.
The left-hand side of inequality (\ref{ckwine}) is the bipartite
entanglement between a probe subsystem $p_1$ and the remaining
subsystems taken as a whole. The right-hand side is the total
bipartite entanglement between $p_1$ and each  of the other
subsystems $p_{j \ne 1}$ in the respective reduced states. Their
difference represents the \textit{residual multipartite
entanglement}, not encoded in pairwise form. For $N=3$, the residual
entanglement quantifies the genuine tripartite entanglement shared
by the three subsystems \cite{CKW,wstates,contangle}. \ineq{ckwine}
is known to hold for spin chains ($N$-qubit systems)
\cite{CKW,osborne} and harmonic lattices ($N$-mode Gaussian states)
\cite{contangle,hiroshima}, with important consequences for the
structure of correlations of those many-body systems
\cite{faziorev,verrucchimany,acinferraro}.

\noindent{\em Is multipartite entanglement monogamous?---} In the
present Letter we wish to investigate if and to what extent sharing
constraints can be established not only for bipartite but also for
multipartite entanglement. In other words, is there a suitable
generalization of the tripartite analysis to arbitrary $N$, such
that a genuine $N$-partite entanglement quantifier is naturally
derived from a stronger monogamy inequality? This question is
motivated by the fact that the residual multipartite entanglement
emerging from the ``weak'' inequality (\ref{ckwine}) includes all
manifestations of $K$-partite entanglement, involving $K$ subsystems
at a time, with $2<K\le N$. Hence, it severely overestimates the
genuine $N$-partite entanglement for $N>3$. It then seems compelling
to further decompose the residual entanglement. How can one
subsystem be entangled with the group of the remaining $N-1$
subsystems? Quite naturally, it can share individual pairwise
entanglement with each of them; and/or genuine three-partite
entanglement involving any two of them (and so on); and/or it can be
genuinely $N$-party entangled with all of them. We then advance the
hypothesis that these contributions are well-defined and mutually
independent, and check {\it a posteriori} that this is indeed true.
Namely, we wish to verify whether in multipartite states,
entanglement is {\it strongly monogamous} in the sense that the
following {\em equality} holds:
\begin{eqnarray}
\label{strongmono}
E^{p_1|(p_2 \ldots p_N)} &=& \begin{array}{c}\sum_{j=2}^N
E^{p_1|p_j} + \sum_{k>j=2}^N E^{p_1|p_j|p_k} \end{array} \nonumber
\\ &+&  \ldots + E^{\underline{p_1}|p_2|\ldots|p_N}\,,
\end{eqnarray}
where $E^{p_1|p_j}$ is the bipartite entanglement between parties
$1$ and $j$, while all the other terms are multipartite
entanglements involving three or more parties. The last contribution
in \eq{strongmono} is defined implicitly by difference and
represents the residual $N$-partite entanglement. It  depends in
general on the probe system $p_1$ with respect to which entanglement
is decomposed. One then needs to define the {\em genuine}
$N$-partite entanglement as the minimum over all the permutations of
the subsystem indexes, $E^{p_1|p_2|\ldots|p_N} \equiv
\min_{\{i_1,\ldots,i_N\}}
E^{\underline{p_{i_1}}|p_{i_2}|\ldots|p_{i_N}}$. All the
multipartite entanglement contributions appearing in \eq{strongmono}
(except the last one) involve $K$ parties, with $K<N$, and are of
the form $E^{p_1|p_2|\ldots|p_K}$. Each of these terms is defined by
\eq{strongmono} when the left-hand-side is the $1 \times (K-1)$
bipartite entanglement $E^{p_1|(p_2 \ldots p_K)}$. The $N$-partite
entanglement is thus, at least in principle, {\em computable} in
terms of the known $K$-partite contributions, once \eq{strongmono}
is applied recursively for all $K=2,\ldots,N-1$. To assess
$E^{p_1|p_2|\ldots|p_N}$ as a proper quantifier of $N$-partite
entanglement, one needs first to show its nonnegativity on all
quantum states. This property in turn implies that \eq{strongmono}
can be recast as a {\em sharper monogamy inequality}, constraining
both bipartite and genuine $K$-partite ($K \le N$) entanglements in
$N$-partite systems. Such a constraint on distributed entanglement
is then a strong generalization of the original CKW inequality
\cite{CKW}, implying it, and reducing to it in the special case
$N=3$.

\begin{figure}[t!]
\includegraphics[width=8.5cm]{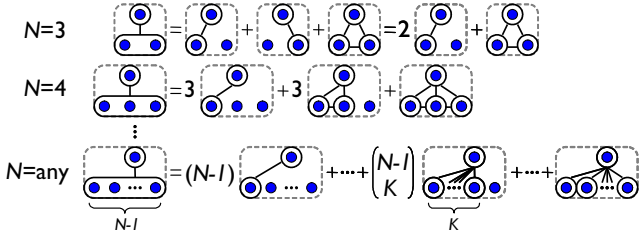}
\caption{(color online)  The structure of multipartite entanglement
in a permutation-invariant state of $N$ parties. The bipartite $1
\times (N-1)$ entanglement is decomposed into all the multiparty
entanglements shared by the single parties. The rightmost graph in
each row depicts the genuine $N$-partite quantum correlations.}
\label{figrafi}
\end{figure}

\noindent{\em Entanglement distribution under permutation
invariance.---} A prominent role in multiparty quantum information
science is played by permutation-invariant (``fully symmetric'')
quantum states.  In practical applications, symmetric states are the
privileged resources for most communication protocols
\cite{definetti}, while from a theoretical perspective they are
basic testgrounds for investigating structural aspects of
multipartite entanglement both for continuous \cite{adescaling} and
discrete \cite{newqubitsnega} variable systems. For our purposes,
specializing to such symmetric states yields a significant
simplification in \eq{strongmono}, as the multipartite entanglements
will only depend on the total number of parties involved in each
contribution. \eq{strongmono} thus reduces to $E^{p_1|(p_2 \ldots
p_N)} = \sum_{K=1}^{N-1} {\binom{N-1}{K} E^{p_1|\ldots|p_{K+1}}}$,
 as pictorially depicted in Fig.~\ref{figrafi}. In
this particular instance, by re-expressing the $K$-partite
contributions in terms of $K'$-partite ones ($\forall \ K'<K$),
the recursion  can be completely resolved, and the proposed
measure of genuine multipartite entanglement in $N$-party
permutation-invariant quantum states takes the following closed
form in terms of a finite sum, with alternating signs, of {\em
bipartite} entanglements:
\begin{equation}\label{ENsym}
E^{p_1|p_2|\ldots|p_N}\!\!=\!\!\!{\begin{array}{c}\sum_{K=1}^{N-1}\!{\binom{N-1}{K}}\end{array}\!\!
(-1)^{K+N+1} E^{p_1|(p_2 \ldots p_{K+1})}}\!.\!
\end{equation}
All the considerations so far are not relying on any specific
 Hilbert space dimensionality.

\noindent{\em Strong monogamy of Gaussian entanglement.---} In the
following, we demonstrate that entanglement indeed distributes
according to the strong monogamy construction \eq{strongmono} in
arbitrary (pure and mixed) $N$-mode Gaussian states on harmonic
lattices, endowed with permutation invariance. We extract a proper,
computable measure of genuine multipartite entanglement by
explicitly evaluating \eq{ENsym}, with $E$ denoting in this case the
 {\em Gaussian contangle} \cite{contangle}. Such $N$-partite entanglement
turns out to be monotone in the optimal fidelity of $N$-party
teleportation networks with symmetric Gaussian resources
\cite{network,telepoppy}, thus acquiring an operational
interpretation and a direct experimental accessibility.

\begin{figure*}[t!]
\includegraphics[width=14cm]{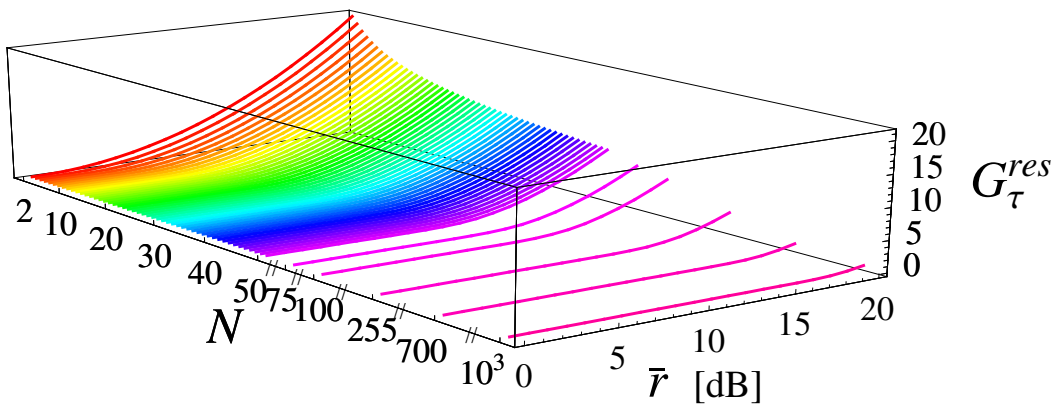} \caption{(color online)
Genuine multipartite entanglement $G_\tau^{res}$ (in ebits) of pure
permutation-invariant $N$-mode Gaussian states $\sig_N$, plotted as
a function of $N$ and of the  squeezing $\bar r$ (in decibels).}
\label{figon}
\end{figure*}

Some preliminaries are in order. We consider a continuous variable
(CV) system consisting of $N$ canonical bosonic modes. Pure, fully
symmetric (permutation-invariant), $N$-mode Gaussian states provide
key resources for essentially all the so-far implemented multiparty
CV quantum information protocols \cite{brareview}. They can be
experimentally prepared by sending a single-mode squeezed state with
squeezing $r_m$ in momentum and $N-1$ single-mode squeezed states
with squeezing $r_p$ in position, through a network of $N-1$
beam-splitters with tuned transmittivities, as detailed in
\cite{network,telepoppy}. Up to local unitaries, such states are
completely specified by the $2N \times 2N$ covariance matrix (CM)
$\sig^{(N)}$ of the second canonical moments, explicitly given in
Ref.~\cite{adescaling}, which is parametrized by the average
squeezing ${\bar r} \equiv (r_m+r_p)/2$. In general, the determinant
of the reduced $K$-mode CM $\sig_K^{(N)}$ of a fully symmetric
$N$-mode pure Gaussian state is given by
$\det\sig_K^{(N)}=[2 K^2-2 N K+2 (N-K) \cosh
   (4 {\bar r}) K+N^2]/N^2$.
The determinant of the CM $\sig$ is related to the purity $\mu$ of a
Gaussian state by $\mu = (\det\sig)^{-1/2}$. In order to evaluate
\eq{ENsym}, we have to compute all the bipartite $1\times K$
entanglements. In general, the bipartite entanglement between $L$
modes and $K$ modes of a fully symmetric Gaussian state can be
concentrated by local unitary operations onto two modes only, a
process known as {\em unitary localizability} \cite{adescaling}.
This effective two-mode Gaussian state is a minimum-uncertainty
mixed state, completely specified by its global (two-mode) and its
two local (single-mode) purities, given by
$[\det\sig_{L+K}^{(N)}]^{-1/2}$, $[\det\sig_{L}^{(N)}]^{-1/2}$, and
$[\det\sig_{K}^{(N)}]^{-1/2}$, respectively. The entire family of
Gaussian entanglement measures, including the Gaussian contangle
$G_\tau$ \cite{contangle}, can be computed in closed form for
minimum-uncertainty mixed states \cite{ordering}. Equipped with
these results, we can analyze the general instance of a mixed,
$N$-mode fully symmetric Gaussian state with CM $\sig_N^{(N+M)}$,
obtained from a pure $(N+M)$-mode one by tracing out $M$ modes. The
genuine $N$-partite entanglement shared by the $N$ individual modes,
according to \eq{ENsym}, acquires the following explicit expression:
\begin{eqnarray}\label{entj}
G_\tau^{res}(\sig_N^{(N+M)})&=&\begin{array}{c}\sum_{j=0}^{N-2}
{\binom{N-1}{j}}\!(-1)^j
f^\tau_j\end{array}\,, \\
f^\tau_j &\equiv& {\rm arcsinh}^2\!\begin{array}{c}
   \left[\frac{2
   \sqrt{N-1-j} \sinh (2
   \bar r)}{\sqrt{M+N} \sqrt{e^{4
   \bar r} (j+M)+N-j}}\right]\end{array}\,, \nonumber
\end{eqnarray}
which depends only on the average squeezing $\bar r$ and the number
of modes $N$ and $M$. \eq{entj} provides a closed, analytical
formula for the genuine $N$-partite Gaussian contangle
$G_\tau^{res}$ of permutation-invariant Gaussian states, as emerging
from the assumed strong monogamy constraint. In Fig.~\ref{figon} we
plot \eq{entj} for pure states ($M=0$) of up to $N=10^{3}$ modes.

It is apparent that the following holds \\
\noindent {\em Theorem.} $G_\tau^{res}(\sig_N^{(N+M)}) \ge 0$.
\\
\noindent {\em Sketch of the proof.} We consider general sums of the
form $F=\sum_{j=0}^{N-2}
\!\!\!{\begin{array}{c}{\binom{N-1}{j}}\end{array}}\!(-1)^j f_j$
where $f_j$ is a decreasing sequence of the integer $j$ with
$f_{N-1}=0$. Positivity (for any $N$) of such sums depends on the
decay rate of $f_j$ with $j$. We choose for comparison a sequence of
the form $\tilde{f}_j(a,b,c)=(N-1-j)/[(N-1)(c+b j^a)]$. The
corresponding alternating sum $\tilde{F}(a,b,c)$ is positive for $a
\le 1$, and specifically $\tilde{F}(1,b,c)=\Gamma(c/b)\Gamma(N-1)/[b
\Gamma(N-1+c/b)]$ where $\Gamma$ is the Euler function. Now,
$f^\tau_j$ in \eq{entj} is bounded both from above and from below by
functions of the form $\tilde{f}_j(1,b,c)$, for any $N$, $M$, $\bar
r$ \cite{namelyproof}.  Hence, $f^\tau_j=O(\tilde{f}_j|a=1)$, which
yields that the corresponding sum, \eq{entj}, is nonnegative as
well. \hfill $\blacksquare$
\\

We have thus demonstrated that multipartite entanglement, once
properly quantified, {\em is strongly} monogamous, in particular in
Gaussian states on permutation-invariant harmonic lattices.
Similarly, one can show that $G_\tau^{res}$ monotonically increases
with the average squeezing $\bar r$, while it decreases with $N$,
eventually becoming identically null in the field limit $N
\rightarrow \infty$. For mixed states, $G_\tau^{res}$ decreases with
the number $M$ of the traced-out modes, i.e. with the mixedness, as
expected. The monotonically increasing dependence of the $N$-partite
entanglement on the squeezing resource $\bar r$, directly yields it
to be an entanglement monotone under Gaussian local operations and
classical communications \cite{ordering,contangle} which preserve
the state symmetry (i.e.~which produce the same local action on
every single mode). The monotonically decreasing dependence on $N$
can be understood as well, since, according to the strong monogamy
decomposition, with increasing number of modes the residual
non-pairwise entanglement can be encoded in so many different
multipartite forms, that the genuine $N$-partite contribution is
actually frustrated.

\noindent{\em Operational connection with teleportation
networks.---} An interesting experimental setting can be
considered, which provides an operational meaning to \eq{entj} as
a {\em bona fide} measure of genuine $N$-partite entanglement.
Permutation-invariant pure Gaussian states ($M=0$) can be
successfully employed as shared resources to implement $N$-party
teleportation networks, where two parties (Alice and Bob) perform
CV teleportation of unknown coherent states, with the assistance
of the other $N-2$ cooperating parties \cite{network}. The optimal
fidelity ${\cal{F}}^{opt}_{N}$ of the process, which quantifies
operationally the shared $N$-partite entanglement, has been
computed in Ref.~\cite{telepoppy}.
In full qualitative and quantitative analogy with $G_\tau^{res}$,
${\cal{F}}^{opt}_{N}$ always lies above the classical threshold
${\cal {F}}^{cl} \equiv 1/2$ (which quantifies the best possible
transfer without using entanglement \cite{network}), it is
monotonically increasing with the squeezing $\bar r$, and
monotonically decreasing with $N$. In fact, for any $N$, the genuine
multipartite entanglement can be recast as a {\em monotonic}
function of ${\cal{F}}^{opt}$, which is explicitly obtained by
substituting  $\bar{r}=\frac14 \log \{1 +
[N(2{\cal{F}}^{opt}-1)]/[2({\cal{F}}^{opt}-1)^2]$ in \eq{entj}
\cite{telepoppy}, with $M=0$. Experimentally, this means that one
does not need a full tomographic reconstruction of the $N$-mode
states to measure $N$-partite entanglement: it can be indirectly
quantified by the success of the teleportation protocol.
Actually, it is enough to measure the quadrature squeezing in any
single mode, and thus $\bar r$, to have a complete information on
{\em any} form of multipartite entanglement of symmetric Gaussian
states. From a broader perspective, the equivalence between
operational entanglement quantifiers (optimal fidelity) and
monogamy-based measures (residual contangle), entails  that there is
 a {\em unique} form of genuine $N$-partite entanglement (for any $N$) in
symmetric Gaussian states, generalizing the results known for
$N=2,3$ \cite{ordering,contangle,telepoppy}.

\noindent{\em Monogamy beyond single modes and promiscuity.---} The
standard monogamy inequalities established so far for spins and
harmonic lattices focus on multipartitions where each subsystem
consists only of one elementary unit (qubit or mode)
\cite{CKW,osborne,contangle,hiroshima}. We will now generalize the
strong monogamy constraint to an arbitrary number of modes per
subsystem. As the unitary localizability of symmetric Gaussian
entanglement applies to general $L \times K$ bipartitions
\cite{adescaling}, \eq{ENsym} can be evaluated explicitly for
subsystems of arbitrary dimension. An important instance is when the
subsystem permutation invariance is preserved: namely, when we
consider a $(n N)$-mode fully symmetric Gaussian state,
multipartitioned in $N$ subsystems, each being a ``molecule'' made
of $n > 1$ modes. In this case,  one immediately sees that
$\det\sig_{n K}^{(n N)}$ does not depend on the integer scale factor
$n$. Therefore, \eq{entj} describes in general the molecular
$N$-partite entanglement in a permutation-invariant $(n N)$-mode
harmonic lattice, which is {\em independent} of the size $n$ of the
molecule: $N$-partite entanglement is, in this sense, {\em scale
invariant}. This is relevant in view of practical exploitation of
Gaussian resources for communication tasks \cite{brareview}:  adding
redundance, e.g. by doubling the size of the individual subsystems,
yields {\em no advantage} for the multiparty-entangled resource.
Importantly, the positivity of \eq{entj} directly entails that
strong monogamy holds as well as a constraint on entanglement
distributed among subsystems formed of arbitrarily many modes, under
permutation invariance. On the other hand, if we keep the number of
modes $N$ fixed, this argument together with the fact that \eq{entj}
decreases with $N$, implies as a general rule that a smaller number
of larger molecules shares strictly more entanglement than a larger
number of smaller molecules.


All forms of $K$-partite entanglement ($2 \le K \le N$) are indeed
{\em simultaneously} coexisting in $N$-partite (pure or mixed)
permutation-invariant Gaussian states and, being the general
expression \eq{entj} an increasing function of $\bar r$, they are
all increasing functions of each other and mutually enhanced. This
structural property of distributed entanglement is known as {\em
promiscuity} and is peculiar to high-dimensional (in the limit,
infinite) spaces \cite{unlim}. In fact, monogamy (already in its
weak form) acts in low-dimensional spaces like those of qubits, such
as to make bipartite and genuine multipartite entanglements mutually
incompatible \cite{wstates}. In Gaussian states of CV systems, full
promiscuity actually occurs under permutation invariance, and is
perfectly compatible with strong monogamy of multipartite
entanglement. This generalizes the known results originally obtained
in permutation-invariant three-mode Gaussian states, which due to
promiscuity have been dubbed the simultaneous analogues of
Greenberger-Horne-Zeilinger and $W$ states of three qubits
\cite{contangle}. Moreover, unlimited promiscuity occurs in a family
of {\em nonsymmetric} four-mode Gaussian states, and strong monogamy
holds as well in that case \cite{unlim}. This fact suggests that the
approach presented here may retain its validity beyond the fully
symmetric scenario.

\noindent{\em Concluding remarks.---} In this Letter we have
addressed and analytically solved the problem of quantifying
genuine multipartite entanglement among (groups of) modes in
Gaussian states on permutation-invariant harmonic lattices. Such
entanglement is experimentally accessible and operationally
related to the fidelity of teleportation networks. The results
obtained for the Gaussian scenario rest on a more general approach
that postulates the existence of stronger monogamy constraints on
distributed bipartite and multipartite entanglement. In this
respect, permutation-invariant states
lend themselves naturally to be investigated via our framework.

Our analysis bears a promising potential in the context of quantum
cryptography: (weak) monogamy of entanglement is the only
requirement that any physical theory must fulfill  to make two-party
quantum key distribution unconditionally secure \cite{monosecure}.
Strong monogamy may likely play the same role as soon as multiparty
secure communication schemes (such as Byzantine agreement, which in
the CV case is solved with fully symmetric Gaussian states
\cite{anna}) are concerned. Further investigation is needed on such
an intriguing topic, as well as on the demonstration of  the strong
monogamy property in other systems, like  $N \geq 4$ qubits,
possibly making use of the techniques introduced in Ref.~\cite{loh}.
\\
\noindent Discussions with
S. De Siena, M. Ericsson, A. Serafini, A. Sanpera, T. Osborne, A. Winter, W. K.
Wootters, M. Piani, and P. Di Gironimo are warmly acknowledged. G. A.
is grateful to G. B. Adesso Jr. for continuous inspiration.




\begin{thebibliography}{99}

\bibitem{hororev}
 R. Horodecki {\it et al.}, arXiv:quant-ph/0702225.


\bibitem{faziorev}
 L. Amico {\it et al.}, arXiv:quant-ph/0703044.


\bibitem{nocloning} W. K. Wootters and W. H. Zurek, Nature {\bf 299}, 802
(1982).

\bibitem{pisa}
B. M. Terhal, IBM J. Res. \& Dev. {\bf 48}, 71 (2004).

\bibitem{CKW}
V. Coffman {\it et al.},
Phys. Rev. A
{\bf 61}, 052306 (2000).

\bibitem{osborne}
T.J.Osborne\,and\,F.Verstraete,\,Phys.Rev.Lett.\,{\bf
96},\,220503\,(2006).

\bibitem{contangle}
G. Adesso and F. Illuminati, New J. Phys. {\bf 8}, 15 (2006).

\bibitem{hiroshima} T. Hiroshima {\it et al.},
Phys. Rev. Lett. {\bf 98}, 050503 (2007).

\bibitem{verrucchimany} T. Roscilde {\it et al.}, Phys. Rev. Lett. \textbf{93}, 167203
(2004).

\bibitem{acinferraro} A. Ferraro {\it et al.},
arXiv:quant-ph/0701009.

\bibitem{wstates} W. D\"ur {\it et al.},
Phys. Rev. A {\bf 62}, 062314 (2000).


\bibitem{definetti} R. Renner, Nature Phys. {\bf 3}, 645 (2007).


\bibitem{adescaling} G. Adesso {\it et al.}
Phys. Rev. Lett. {\bf 93}, 220504 (2004); A. Serafini {\it et al.},
Phys. Rev. A {\bf 71}, 032349 (2005).

\bibitem{newqubitsnega} A. R. U. Devi {\it et al.},
Phys. Rev. Lett. {\bf 98}, 060501 (2007).


\bibitem{network}
P.\,van\,Loock\,and\,S.L.Braunstein,\,Phys.Rev.Lett.\,{\bf
84},\,3482\,(2000).

\bibitem{telepoppy}
G. Adesso and F. Illuminati, Phys. Rev. Lett. {\bf 95}, 150503
(2005).

\bibitem{brareview} S. L. Braunstein and P. van Loock, Rev. Mod. Phys. {\bf 77}, 513
(2005); G. Adesso and F. Illuminati, J. Phys. A {\bf 40}, 7821 (2007).



\bibitem{ordering}
G. Adesso, and F. Illuminati, Phys. Rev. A {\bf 72}, 032334 (2005).

\bibitem{namelyproof}
Explicitly, $\tilde f(1,b_0,c_0) \le f^\tau_j \le f(1,1,c_0)$, with
$c_0 = 1/{\rm arcsinh}^2 \{[2
   \sqrt{N-1} \sinh (2
   \bar r)]/[\sqrt{M+N} \sqrt{e^{4 \bar r}
   M+N}]\}$ and $b_0=
\{[2
   (e^{4 \bar r}
   (M+N-1)+1) \sinh (2
   \bar r)]/[(e^{4 \bar r} M+N)
   \sqrt{(N-1) (4 (N-1)
   \sinh ^2(2 \bar r)+(M+N)
   (e^{4 \bar r}
   M+N))}]-1/[(N-1)
   \sqrt{c_0}]\}
   c_0^{3/2}$.


\bibitem{unlim} G. Adesso {\it et al.},
Phys. Rev. A {\bf 76}, 022315 (2007).



\bibitem{monosecure} M. Pawlowski, arXiv:0705.2155.

\bibitem{anna}
R. Neigovzen and A. Sanpera, arXiv:quant-ph/0507249.


\bibitem{loh}
R. Lohmayer {\it et al.}, Phys. Rev. Lett. {\bf 97}, 260502 (2006);
J.-M. Cai {\it et al.}, Phys. Lett. A {\bf 363}, 392 (2007).



\end{thebibliography}
\end{document}